\documentclass[12pt,smallextended,natbib,runningheads,epsfig]{article}
\usepackage{latexsym, amsfonts}
\usepackage{amssymb, amsmath}
\usepackage{amsfonts}
\usepackage{amsfonts}
\usepackage{graphicx}
\usepackage{adjustbox}
\usepackage{setspace}
\usepackage{rotating}
\usepackage{tikz}
\usepackage{bigints}
\usepackage{bm}
\usepackage{multirow}
\usepackage{subcaption}
\usepackage{float}
\usepackage{changepage}  
\newcommand{\be}{\begin{equation}}
\newcommand{\ee}{\end{equation}}
\newcommand{\bea}{\begin{eqnarray}}
\newcommand{\eea}{\end{eqnarray}}
\newcommand{\bean}{\begin{eqnarray*}}
\newcommand{\eean}{\end{eqnarray*}}
\newcommand{\brray}{\begin{array}}
\newcommand{\erray}{\end{array}}
\newcommand{\ben}{\begin{equation}{nonumber}}
\newcommand{\een}{\end{equation}{nonumber}}

\newtheorem{dfn}{Definition}[section]
\newtheorem{thm}[dfn]{Theorem}
\newtheorem{lmma}[dfn]{Lemma}
\newtheorem{ppsn}[dfn]{Proposition}
\newtheorem{crlre}[dfn]{Corollary}
\newtheorem{xmpl}[dfn]{Example}
\newtheorem{rmrk}[dfn]{Remark}

\newcommand{\bdfn}{\begin{dfn}}
\newcommand{\bthm}{\begin{thm}}
\newcommand{\blmma}{\begin{lmma}}
\newcommand{\bppsn}{\begin{ppsn}}
\newcommand{\bcrlre}{\begin{crlre}}
\newcommand{\bxmpl}{\begin{xmpl}}
\newcommand{\brmrk}{\begin{rmrk}}
\newcommand{\edfn}{\end{dfn}}
\newcommand{\ethm}{\end{thm}}
\newcommand{\elmma}{\end{lmma}}
\newcommand{\eppsn}{\end{ppsn}}
\newcommand{\ecrlre}{\end{crlre}}
\newcommand{\exmpl}{\end{xmpl}}
\newcommand{\ermrk}{\end{rmrk}}

\def\a*{{\cal A}_{h,*}}
\def\B{{\cal B}(h)}
\def\B1{{\cal B}_1(h)}
\def\b{{\cal B}^{\rm s.a.}(h)}
\def\b1{{\cal B}^{\rm s.a.}_1(h)}

\numberwithin{equation}{section}

\begin{document}
\begin{center}
{\Large {\bf Bayesian Modeling of Long-Term Dynamics in Indian Temperature Extremes}}\\[1.5ex]
{\large Chitradipa Chakraborty\textsuperscript{*}}\\
\end{center}

\begin{adjustwidth}{-0.5cm}{-0.5cm} 
\begin{abstract}
Annual maximum temperature data provides crucial insights into the impacts of climate change, especially for regions like India, where temperature variations have significant implications for agriculture, health, and infrastructure. In this study, we propose the Coupled Continuous Time Random Walk (CTRW) model to analyze annual maximum temperature data in India from 1901 to 2017 and compare its performance with the Bayesian Spectral Analysis Regression (BSAR) model. The CTRW model extends the standard framework by coupling temperature changes (jumps) and waiting times, capturing complex dynamics such as memory effects and non-Markovian behavior. The BSAR model, in contrast, combines a linear trend component with a non-linear isotonic function, modeled using a Gaussian Process (GP) prior, to account for smooth and flexible non-linear variations in temperature. By applying both models to the temperature data, we evaluate their ability to capture long-term trends and seasonal fluctuations, offering valuable insights into the effects of climate change on temperature dynamics in India. The comparison highlights the strengths and limitations of each approach in modeling temperature extremes and provides a robust framework for understanding climate variability.

\vspace{0.2 in}

\end{abstract}
\end{adjustwidth}

\renewcommand{\thefootnote}{\fnsymbol{footnote}} 
\footnotetext{\hspace{-0.5em}$^{*}$Beijing Key Laboratory of Topological Statistics and Applications for Complex Systems, Beijing Institute of Mathematical Sciences and Applications, Beijing 101408, China.}
\footnotetext{\hspace{-0.5em}Email: \hspace{-0.5em} \texttt{chitradipachakraborty@gmail.com}}
\vspace{0.1 in}

\section{Introduction}
In recent years, there has been a strong and growing scientific consensus that the Earth is experiencing a significant and sustained warming trend. According to the Intergovernmental Panel on Climate Change (IPCC) reports, global temperatures have been rising steadily since 1850, with the past decade recorded as the warmest in history. A key driver behind this phenomenon is the rapid increase in atmospheric carbon dioxide (CO$_2$) levels, predominantly resulting from human activities such as the burning of fossil fuels, deforestation, and industrial processes (Le Quéré et al., 2018; Friedlingstein et al., 2020). Elevated CO$_2$ concentrations intensify the greenhouse effect, whereby heat emitted from the Earth’s surface is trapped in the atmosphere, leading to higher global temperatures (Held \& Soden, 2006). The consequences of global warming are expected to be profound and far-reaching. Increased CO$_2$ not only raises atmospheric temperature but also contributes to ocean acidification, threatening marine ecosystems and leading to the potential collapse of coral reefs and other aquatic biodiversity (Doney et al., 2009). Furthermore, global warming is associated with more frequent and severe extreme weather events, including hurricanes, droughts, floods, and wildfires (Pizzorni et al., 2024; Bolan et al., 2024). Rising sea levels, driven by the melting of glaciers and polar ice caps, pose additional threats to coastal communities (Church et al., 2013). Climate change is also anticipated to trigger mass migrations, food insecurity, and health crises as natural systems become increasingly disrupted (Trummer et al., 2023).\\
In this context, it is crucial to develop accurate and interpretable models that can track and predict temperature changes over time. The present study focuses on modeling long-term temperature trends using the All India Annual Maximum Temperature Series (1901–2017), which exhibits clear signs of warming in recent decades. Bayesian methodologies provide a coherent framework for modeling uncertainty, incorporating prior knowledge, and drawing probabilistic inferences, which are particularly valuable in climate studies characterized by sparse, noisy, or irregularly sampled data. In particular, semiparametric Bayesian models allow for flexible estimation of nonlinear trends while retaining interpretability, building on foundational contributions in Bayesian robustness and model comparison (Delampady \& Ghosh, 1999), the importance of predictive model assessment (Delampady, 2001), and the judicious choice of priors in complex hierarchical frameworks. Inspired by these perspectives, we adopt a Bayesian approach to analyze the dynamic behavior of temperature extremes and evaluate competing models using model selection criteria.\\
To analyze this data, we propose a coupled Continuous Time Random Walk (CTRW) model (Meerschaert \& Scheffler, 2010) within a Bayesian framework, where the jump length is correlated with the waiting time. This approach allows for flexible modeling of irregular, stochastic changes in temperature over time. 
The application of this framework to annual maximum temperature modeling is motivated by the episodic and regime-like behavior frequently observed in the Indian climate system. Though temperature is often treated as a continuous process, empirical evidence suggests that temperature extremes can exhibit abrupt shifts followed by prolonged stagnation, reflecting the influence of large-scale climate phenomena such as El Niño–Southern Oscillation or ENSO (Cai et al., 2021), delayed monsoon withdrawal (Goswami et al., 2006; Kothawale et al., 2010), and aerosol-induced cooling (Chen et al., 2024). These features closely align with the jump–wait behavior captured by the coupled CTRW framework, which explicitly accounts for memory effects and non-Markovian dynamics (Zaburdaev et al., 2015). This makes coupled CTRW particularly well-suited for modeling the stochastic and irregular nature of climate variations in the Indian context.\\
To evaluate the performance of our proposed model, we compare it with the Bayesian Spectral Analysis Regression (BSAR) model (Lenk \& Choi, 2017), which employs a partially linear semiparametric regression between time and temperature, incorporating an unknown isotonic (non-decreasing) function governed by a Gaussian process prior. This comparative analysis aims to assess the suitability and predictive power of the coupled CTRW model in capturing the evolution of temperature patterns in India and, ultimately, to provide insights that may inform climate policy and adaptation strategies.\\
The remainder of the paper is organized as follows. Section 2 provides a detailed description of the All India Annual Maximum Temperature dataset, outlining its temporal span and the key variables considered in this analysis. Section 3 presents the methodology, starting with the mathematical formulation and assumptions behind the coupled CTRW model, followed by the BSAR with isotonic Gaussian Process model and its application to the temperature data. In Section 4, we present the results of applying both models to the temperature series, comparing their performance in terms of fitting accuracy and interpretability, and highlighting their respective strengths and limitations. Finally, Section 5 discusses the implications of the key findings and offers directions for future research.

\section{The Maximum Temperature Data}
The dataset used in this study is the All India Annual Maximum Temperature Series, spanning the period from 1901 to 2017. It was released under the National Data Sharing and Accessibility Policy (NDSAP) and contributed by the Ministry of Earth Sciences (MoES) and the India Meteorological Department (IMD). This dataset falls under the thematic sectors of Earth Sciences, Atmospheric Science, and Science \& Technology and was officially published on the Indian government's open data portal on September 23, 2016. \\ 
The dataset includes yearly maximum temperature measurements averaged across India, serving as a nationally aggregated indicator of extreme temperature conditions. Each entry in the annual series corresponds to the highest recorded temperature (in degree Celsius) for a given calendar year. In addition to the annual values, the dataset also contains seasonal maximum temperatures specifically, the highest temperatures recorded within each of the four climatological seasons: Winter (January–February), Pre-monsoon (March–May), Monsoon (June–September), and Post-monsoon (October–December). These seasonal maxima offer additional insights into intra-annual variation and allow for a finer-grained analysis of\\

\begin{figure}[htbp]
    \centering
    \includegraphics[width=0.9\textwidth]{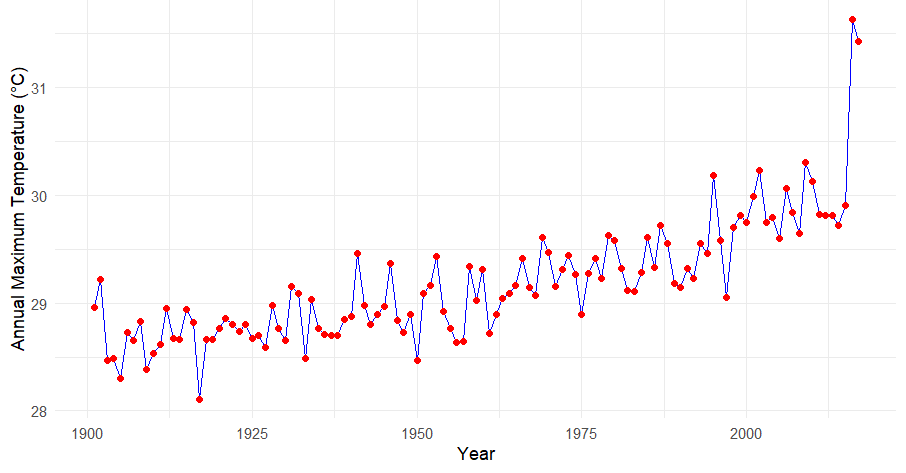}
    \caption{Annual Maximum Temperature Trend in India (1901–2017)}
    \label{fig:temp_trend}
\end{figure}

\noindent  temperature extremes. The long temporal span of 117 years makes this dataset particularly suitable for trend analysis, change-point detection, and the modeling of long-term temperature dynamics (Rajeevan et al., 2010; Dash \& Hunt, 2007). Figure 1 shows the temporal trend of annual maximum temperatures in India from 1901 to 2017. \\

\begin{figure}[htbp]
    \centering
    \includegraphics[width=0.9\textwidth]{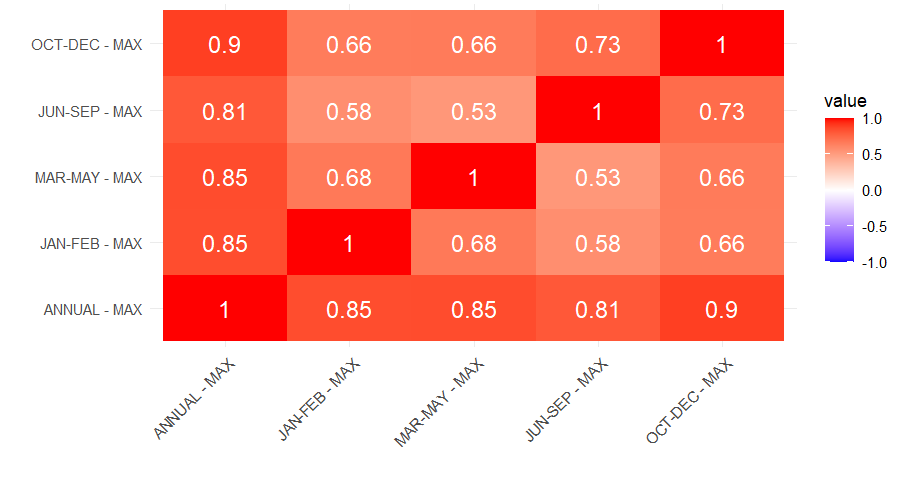}
    \caption{Correlation Matrix between Annual and Seasonal Maximum Temperatures}
    \label{fig:corr_mat}
\end{figure} 

\noindent Figure 2 displays the correlation matrix between annual and seasonal maximum temperatures. Notably, the maximum temperature during the Post-monsoon season exhibits the highest correlation (0.90) with the annual maximum temperature. In many parts of India and South Asia, October can experience unusually high temperatures due to clear skies following monsoon withdrawal, trapped humidity and heat, and delayed retreat of hot air masses (Sikka \& Gadgil, 1980). In recent decades, climate change has contributed to a warming trend extending into late autumn and early winter, resulting in record-setting warm days during the October–December period, sometimes exceeding the historically typical “peak summer” months. As a result, peak annual maximum temperatures are increasingly being observed during the Post-monsoon season, reflecting evolving climate patterns and delayed monsoon retreat. This shift in temperature patterns makes the dataset particularly suitable as a robust empirical foundation for applying advanced statistical models, such as the Coupled Continuous Time Random Walk (CTRW) model and the Bayesian Spectral Analysis Regression (BSAR) with Isotonic Gaussian Process model, to explore the underlying dynamics of climate change in the Indian context.

\section{The Methodology}
\subsection{The Coupled CTRW Model}
A Coupled Continuous Time Random Walk (CTRW) is an extension of the classical CTRW framework introduced by Montroll \& Weiss (1965), where jump lengths and waiting times are coupled (Metzler \& Klafter, 2000; Zaburdaev, 2006). This coupling enables the model to capture complex dynamics observed in real-world systems, such as anomalous diffusion, non-Markovian behavior, and memory effects (Pu et al., 2024; Scalas et al., 2000; Magdziarz et al., 2009), which are often observed in environmental and climate processes.\\ 
In the context of modeling annual maximum temperature, the coupled CTRW model is specified as: $$T_i = T_{i-1} + \delta_i \cdot w_i,$$ where $T_i$ denotes the annual maximum temperature in year $i$, $\delta_i$ is the jump size representing the temperature change, and $w_i$ is the waiting time between jumps. The term $\delta_i \cdot w_i$ reflects the coupling between the jump size and the waiting time, implying that the magnitude of the temperature change is influenced by the duration between events. For instance, large temperature changes might occur after long waiting times, or possibly short waiting times, depending on the nature of the coupling. \\

\begin{figure}[htbp]
    \centering
    \includegraphics[width=0.9\textwidth]{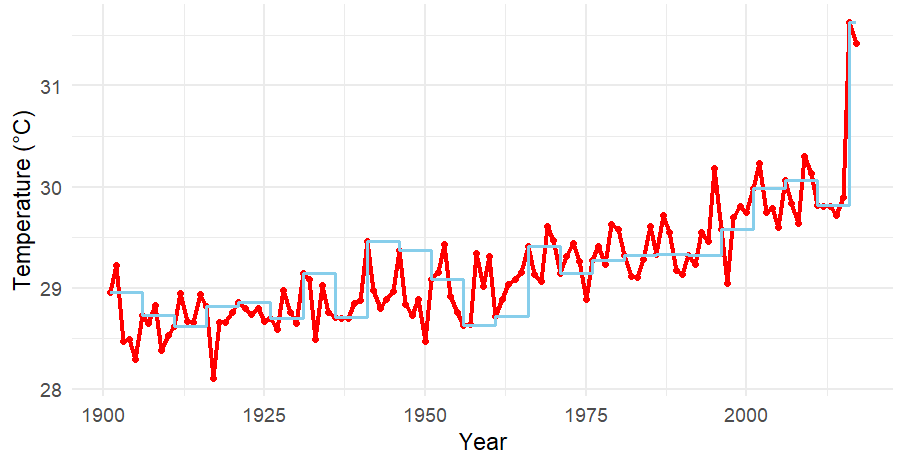}
    \caption{Schematic View on a Continuous Time Random Walk}
    \label{fig:ctrw_step}
\end{figure}

\noindent The model assumes that the jump sizes $\delta_i$ follow a normal distribution, i.e., $\delta_i \sim N(\mu, \tau^2)$ and the waiting times $w_i$ follow a Gamma distribution, i.e., $w_i \sim \text{Gamma}(\alpha, \beta)$. Although the distributions of $\delta_i$ and $w_i$ are assumed to be independent, their product $\delta_i \cdot w_i$ introduces non-linear memory into the model. This coupling means that the temperature change at time $i$ depends not only on the current state but also on the past waiting times and jump sizes, capturing long-term dependencies in the temperature dynamics. The resulting product distribution of $\delta_i \cdot w_i$ can exhibit skewness and heavy tails, reflecting the complex, memory-driven behavior that is often observed in climate systems.

\subsection{The BSAR with Isotonic Gaussian Process Model}

The Bayesian Spectral Analysis Regression (BSAR) model, as introduced by Lenk \& Choi (2017), combines both linear and non-linear components to effectively model time series data. The linear component captures the long-term trend, while the non-linear component, modeled using a Gaussian Process (GP) prior, captures smooth and flexible non-linear deviations from linearity (Rasmussen \& Williams, 2006). This hybrid approach is particularly suitable for systems that exhibit long-term, smoothly increasing trends, such as global temperature rise.\\
In the context of modeling annual maximum temperature, the BSAR model is specified as:
\[
T_i = \beta_0 + \beta_1 X_i + f(t_i) + \epsilon_i,
\]
where $T_i$ represents the annual maximum temperature in year $i$, $\beta_0$ and $\beta_1$ are the regression coefficients, $X_i$ denotes the year, and $t_i \in \{1, 2, \dots, N\}$ is the time index used for evaluating the isotonic Gaussian process function $f(t_i)$. The function $f \in \mathcal{F}^+$ is an unknown isotonic regression function (Dette et al., 2006) with $\mathcal{F}^+ = \left\{ f \ | \ f(t_1) \leq f(t_2) \ \forall \ t_1 < t_2 \right\}$, denoting the space of non-decreasing functions (Lin \& Dunson, 2014). To impose monotonicity, the derivative of $f$ is modeled as the square of a Gaussian process (GP), guaranteeing non-negativity. According to Lenk \& Choi (2017), the function $f(x)$ can be written as
\[
f(x) = \gamma^2 \left( \int_0^x Z^2(s) \, ds - \int_0^1 \int_0^x Z^2(s) \, ds \, dx \right),
\]
where $Z$ is a second-order Gaussian process with zero mean and covariance function $\nu(s, t) = \mathbb{E}[Z(s)Z(t)]$ for $s, t \in [0, 1]$, and $\gamma^2$ is a smoothing parameter controlling the shape of $f$. The residual term is assumed to follow a normal distribution, $\epsilon_i \sim N(0, \sigma^2)$. The monotonicity assumption imposed through the squared GP derivative is aligned with physical expectations of long-term global warming trends. This makes the BSAR model well-suited to capturing persistent upward trends in annual temperature extremes, while still allowing for non-linear flexibility. 

\section{The Analysis}

This section presents the results of fitting both the coupled CTRW model and the BSAR with isotonic Gaussian Process model to the annual maximum temperature data. Both models were estimated within a Bayesian framework using Markov Chain Monte Carlo (MCMC) methods implemented in JAGS (Plummer, 2003). Each model was run with four parallel chains of 10,000 iterations, with the first 2,000 discarded as burn-in. Thinning was applied every 10 iterations to reduce autocorrelation (Gelman et al., 2013). Convergence was assessed using standard diagnostics such as trace plots and the Gelman-Rubin statistic ($\hat{R}$), with all parameters achieving $\hat{R}$ values close to 1, indicating good convergence (Gelman \& Rubin, 1992). Non-informative or weakly informative priors were employed for most model parameters (Gelman et al., 2008). 
\subsection{Coupled CTRW Model Results}
The coupled CTRW model yielded posterior estimates for key parameters, including the mean and variance of jump lengths ($\mu$, $\tau^2$), and the waiting time distribution parameters ($\alpha$, $\beta$). These parameters govern the frequency and magnitude of temperature changes, capturing the underlying jump–wait dynamics inherent in the annual maximum temperature series (Jurlewicz et al., 2012; Barkai, 2001; Sokolov et al., 2002). The fitted model, as shown in Figure 4, effectively captures phases of rapid temperature increases interspersed with periods of stagnation. Phases of stagnation are evident between 1900 and around 1975,\\

\begin{figure}[H]
    \centering
        \includegraphics[width=0.9\linewidth]{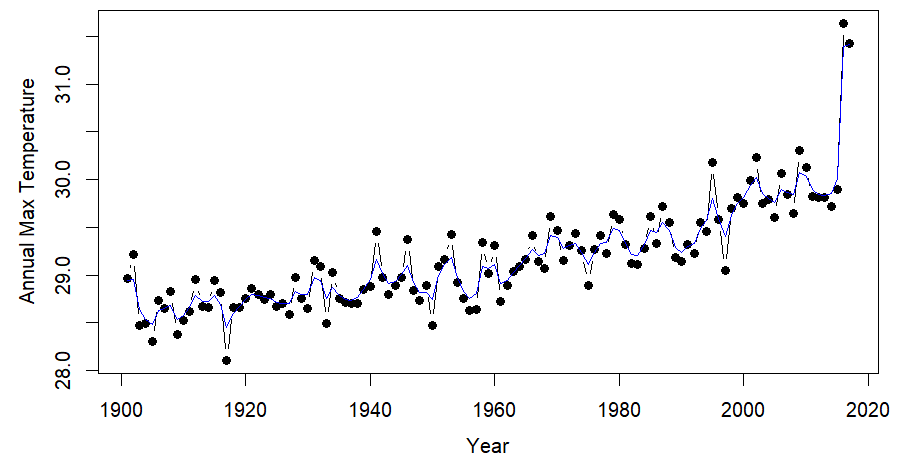}
        \caption{Overlay of Observed Annual Maximum Temperatures (black) and Coupled CTRW Model Predictions (blue)}
        \label{fig:ctrw_fit}
 \end{figure}

\begin{figure}[H]
    \centering
        \includegraphics[width=0.9\linewidth]{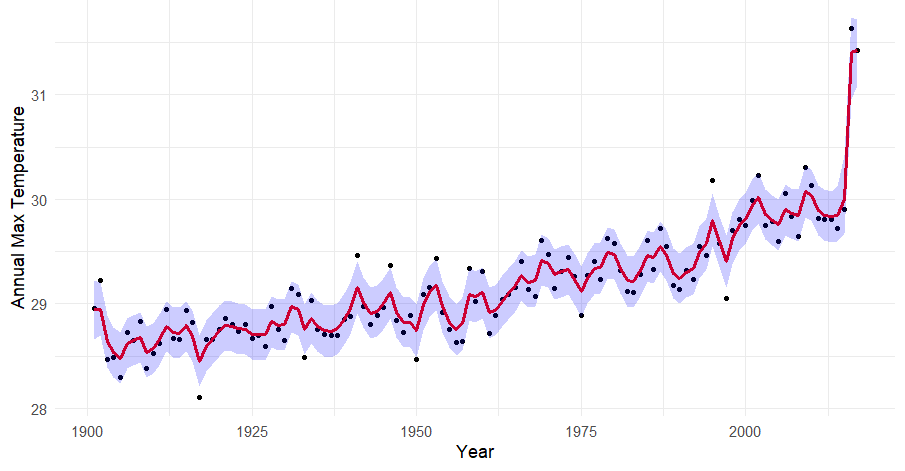}
        \caption{Posterior Predictive Fit of the Annual Maximum Temperature Series using the Coupled CTRW Model, with 95\% Credible Bands}
        \label{fig:ctrw_model}
 \end{figure}

\begin{figure}[H]
    \centering
        \includegraphics[width=0.9\linewidth]{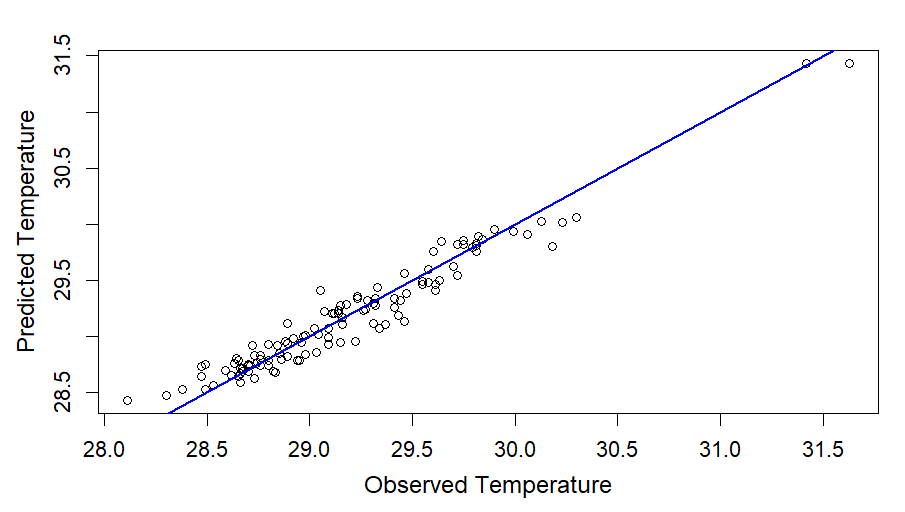}
        \caption{Observed vs Predicted Annual Maximum Temperatures based on the Coupled CTRW Model}
        \label{fig:ctrw_op}
\end{figure}

\noindent where the annual maximum temperature exhibits frequent fluctuations but lacks a strong upward or downward trend, consistent with the waiting times component of CTRW. Additional short stagnation periods appear around 1980–1995 and 2005–2010. In contrast, phases of rapid increase are marked by notable jumps: around 1940–1945, between 1995–2005, and a particularly sharp rise around 2016–2017. This pattern reflects a realistic jump–wait mechanism in the evolution of annual maximum temperatures, as characterized by the coupled CTRW framework. Figure 5 illustrates the coupled CTRW model fit (in red) to the observed temperature series (black dots), with the shaded blue region denoting the 95\% credible interval. Finally, Figure 6 shows a scatter plot comparing observed and predicted annual maximum temperatures, where the tight alignment around the 45-degree line suggests strong predictive accuracy. These results confirm that the coupled CTRW model offers a flexible and interpretable framework for capturing the episodic nature of temperature changes, especially during periods characterized by sudden shifts followed by relative stability, which aligns with episodic heatwaves and pauses likely due to large-scale climatic mechanisms (Metzler \& Klafter, 2004).

\subsection{BSAR with Isotonic GP Model Results}
The BSAR model produced posterior estimates for the linear coefficients ($\beta_0$, $\beta_1$), random error variance $\sigma^2$, and smoothing parameter $\gamma^2$. The use of an isotonic Gaussian Process (GP) prior enabled the model to flexibly estimate a non-decreasing temperature trend over time, without imposing a strict functional form (Choi \& Lenk, 2020; Choudhuri et al., 2004). The isotonic GP utilized a squared exponential kernel, with hyperpriors placed on both the length-scale and variance parameters. This nonparametric approach was particularly effective for capturing gradual long-term warming trajectories while maintaining monotonicity. The fitted model, as shown in Figure 7, captures a smooth and upward trend consistent with the long-term warming pattern observed in the Indian subcontinent. Figure 8 displays the BSAR with isotonic GP model’s posterior fit (in red) superimposed on the observed annual maximum temperature series (black dots), with 95\% credible band (shaded blue). Finally, Figure 9 presents the observed versus predicted values, showing strong overall agreement. Altogether, the BSAR model provides a smoother fit to the temperature series and is particularly effective at capturing gradual, long-term warming trends. Its isotonic constraint ensures monotonicity, aligning well with the overall increase in temperature. However, due to the model’s smoothness assumption (Mammen, 1991), it may struggle to fully capture abrupt regime shifts or short-term variations in temperature dynamics. For example, the BSAR model tends to underrepresent the sharp rise in temperatures observed during the period from 2016 to 2017. This limitation arises from the \\

\begin{figure}[H]
    \centering
        \includegraphics[width=0.9\linewidth]{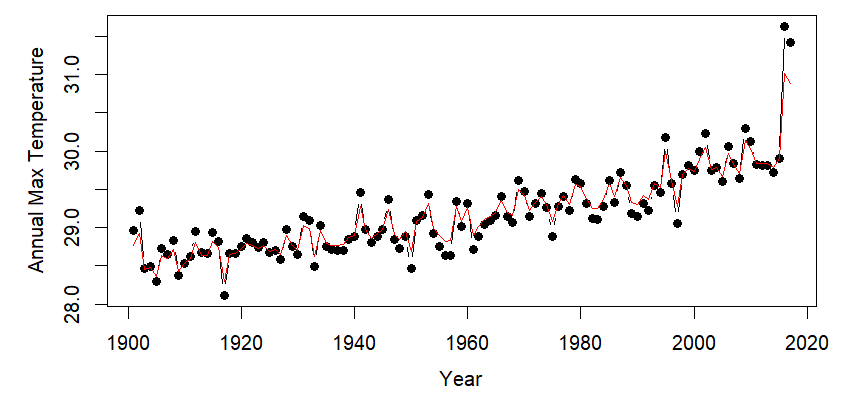}
        \caption{Overlay of Observed Annual Maximum Temperatures (black) and BSAR with Isotonic GP Model Predictions (red)}
        \label{fig:bsar_fit}
\end{figure}
\begin{figure}[H]
    \centering
        \includegraphics[width=0.9\linewidth]{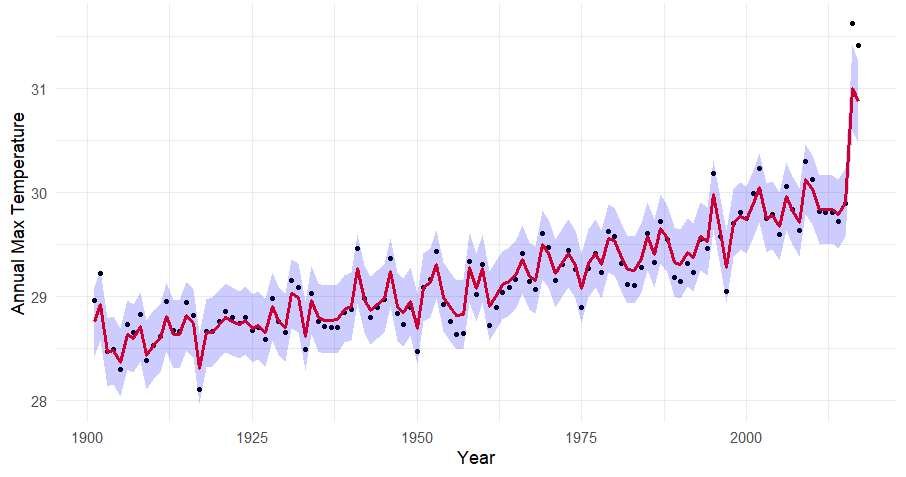}
        \caption{Posterior Predictive Fit of the Annual Maximum Temperature Series using the BSAR with Isotonic GP Model, with 95\% Credible Bands}
        \label{fig:bsar_model}
\end{figure}

\begin{figure}[H]
    \centering
        \includegraphics[width=0.9\linewidth]{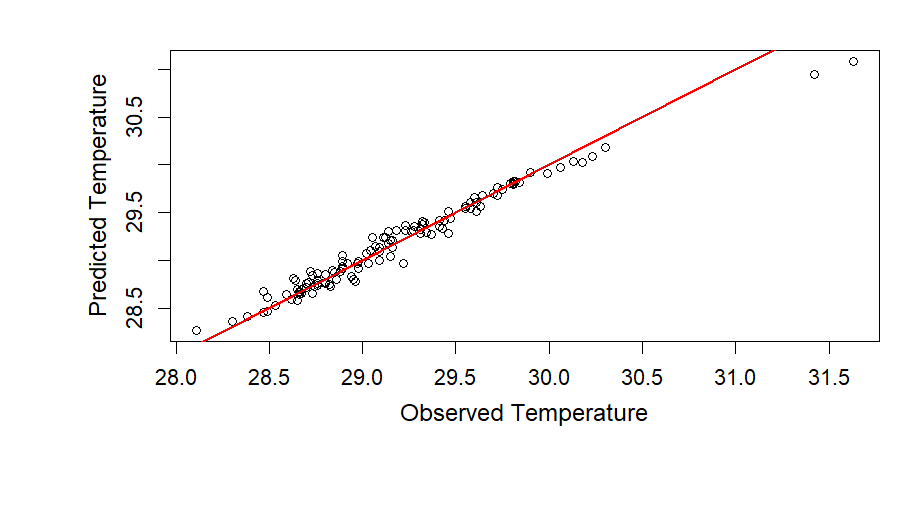}
        \caption{Observed vs Predicted Annual Maximum Temperatures based on the BSAR with Isotonic GP Model}
        \label{fig:bsar_op}
\end{figure}
\noindent model's inherent preference for smooth continuity, highlighting a trade-off between accurately capturing long-term trends and its sensitivity to abrupt transitions.

\subsection{Model Comparison}
The comparative performance of the coupled CTRW and BSAR with isotonic GP models was evaluated using model selection criteria including the Deviance Information Criterion (DIC) (Spiegelhalter et al., 2002), Root Mean Squared Error (RMSE), Mean Absolute Error (MAE), and Leave-One-Out Information Criterion (LOOIC) (Vehtari et al., 2017). From Table 1, the coupled CTRW model is strongly favored in terms of overall model fit (DIC) and generalizability (LOOIC), indicating that it better captures the underlying temperature dynamics across varying temporal regimes. While the BSAR with isotonic GP model shows lower RMSE and MAE, suggesting better in-sample prediction accuracy, these metrics do not account for model complexity or the potential for overfitting. In contrast,\\

\begin{table}[htbp]
    \centering
    \begin{tabular}{|l|c|c|c|c|c|}
        \hline
        \textbf{Model} & \textbf{DIC} & \textbf{RMSE} & \textbf{MAE} & \textbf{LOOIC}  & \textbf{Fit Description} \\
        \hline
        CTRW & 110.048 & 0.134 & 0.105 & 24.009 & Dynamic (jump + waiting) \\
        BSAR & 443.754 & 0.123 & 0.084 & 68.966 & Semiparametric GP + linear \\
        \hline
    \end{tabular}
     \caption{Model Performance Metrics for Coupled CTRW vs BSAR with Isotonic GP}
    \label{tab:model_comparison}
\end{table}

\noindent DIC and LOOIC provide a more comprehensive evaluation, considering the effective number of parameters and assessing predictive performance on held-out data. This divergence in performance metrics highlights an important trade-off: the BSAR model excels in fitting smooth long-term trends but may struggle to generalize in the presence of abrupt changes\\

\begin{figure}[H]
    \centering
    \includegraphics[width=0.95\textwidth]{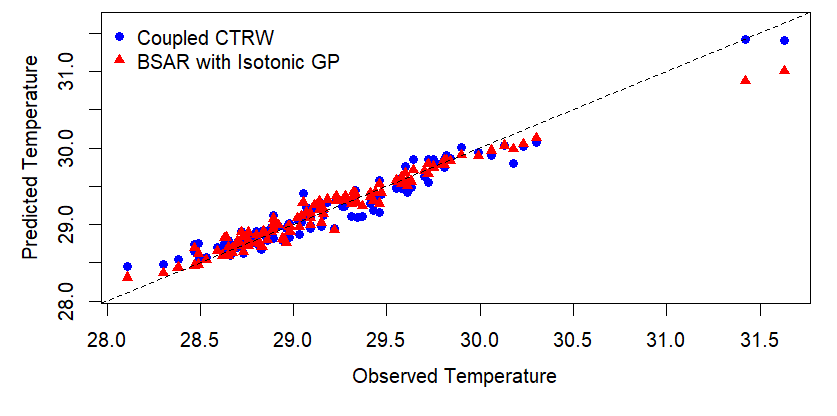}
    \caption{Comparison of Observed vs Predicted Annual Maximum Temperatures using the Coupled CTRW and the BSAR with Isotonic GP}
    \label{fig:ctrw_bsar}
\end{figure}

\noindent  or regime shifts. On the other hand, the CTRW model, with its stochastic burst–pause structure, is better equipped to generalize to irregular patterns, even at the cost of slightly higher pointwise prediction errors.\\
Figure 10 further illustrates model performance by plotting observed versus predicted annual maximum temperatures. The BSAR with isotonic GP model (red triangles) aligns more closely with the observed values, especially in terms of capturing the long-term trend, whereas the coupled CTRW model (blue circles) captures greater variability, potentially linked to temporal irregularities. The dashed line represents the 45-degree line, indicating perfect prediction accuracy. Overall, the results demonstrate that the coupled CTRW model captures anomalous diffusion and memory effects in temperature changes, making it suitable for modeling irregular, stochastic climate variations. In contrast, the BSAR model, with its isotonic GP prior, effectively captures long-term warming trends while allowing for non-linear, monotonic increases in temperature. This comparative analysis highlights the strengths of both models, enabling the selection of the most appropriate model for different climate scenarios: coupled CTRW for modeling irregular fluctuations and BSAR with isotonic GP for smooth, long-term trend estimation.

\section{Conclusion}
This study set out to model the long-term dynamics of annual maximum temperatures in India using two advanced Bayesian frameworks: a Coupled Continuous Time Random Walk (CTRW) model and the Bayesian Spectral Analysis Regression (BSAR) model with an isotonic Gaussian Process prior. Both models were designed to capture the non-linear, non-stationary, and complex nature of temperature extremes over more than a century of observations (1901–2017), accounting for stochastic variations and long-term trends that cannot be explained by simple linear models.\\
The coupled CTRW model, by introducing coupling between the magnitude of temperature changes (jumps) and the time between those changes (waiting times), successfully reproduced the episodic nature of temperature evolution (Shlesinger et al., 1987). The model highlighted distinct periods of rapid temperature increases followed by relative stability, which may correspond to global climate events or regional climatic drivers. Its flexibility in capturing such stochastic burst–pause dynamics makes it particularly well-suited for applications where memory effects and anomalous diffusion-like behaviors are evident.\\
In contrast, the BSAR model provided a complementary perspective. By enforcing an isotonic (non-decreasing) constraint on the smooth trend function via a squared Gaussian Process derivative, the BSAR approach effectively modeled the persistent upward trajectory of temperature over time. It also offered a clearer quantification of the long-term warming trend, without assuming a specific functional form, thereby preserving flexibility while respecting the physical intuition of monotonic climate change.\\
Both models showed strong predictive performance, as evidenced by posterior predictive fits and high correspondence between observed and predicted values. However, they differ in interpretability and applicability. The coupled CTRW model excels in capturing irregular but significant temperature shifts through its jump–wait formalism, while the BSAR with isotonic GP model is more interpretable in terms of trend estimation, particularly when monotonicity is assumed. These contrasting modeling philosophies can thus serve different purposes — CTRW for exploring abrupt, episodic temperature dynamics and BSAR for estimating smooth long-term warming trajectories.\\
From a policy perspective, the insights from both models reinforce the growing concern of escalating temperature extremes in India. The fact that post-monsoon temperatures now exhibit the strongest correlation with annual maxima reflects a shift in the seasonal behavior of heatwaves and global warming (Mishra et al., 2014), which has implications for agriculture, health, and energy planning. Climate adaptation strategies must therefore account not only for mean temperature increases but also for changing temporal patterns and the increased frequency of extreme events.\\
There are some limitations in the current coupled CTRW and BSAR models. The current coupled CTRW specification assumes stationary distributions for jump sizes and waiting times across the full 117-year span. While this simplification enables tractable inference, it may limit the model's ability to reflect the accelerating or evolving nature of climate change. Similarly, due to the smoothness properties of the Gaussian Process and the isotonic constraint, the BSAR model may oversmooth abrupt transitions, potentially failing to capture short-term fluctuations or rapid regime shifts that appear in the observed series (e.g., sharp increases in 2016–2017). The model’s performance is also influenced by the smoothing parameter $\gamma^2$, which governs the flexibility of the isotonic function. While $\gamma^2$ is estimated in a Bayesian framework, its posterior behavior may vary across periods, especially in regions of higher volatility. \\
Several avenues remain open for further exploration. First, exploring time-varying CTRW parameters or regime-switching extensions, allowing the model to adapt to structural changes in temperature dynamics over time (Rosen et al., 2012). Second, extending the CTRW and BSAR models to include external covariates (e.g., CO$_2$ emissions, aerosol concentrations, urbanization indices) can help attribute observed temperature dynamics to anthropogenic and natural factors. Third, adapting the frameworks to a spatiotemporal setting, for example, using spatial Gaussian Processes or Bayesian hierarchical models could enable more nuanced, region-specific climate assessments (Banerjee et al., 2014). Finally, analyzing seasonal maximum temperatures in a multivariate Bayesian framework could offer deeper insight into the evolving intra-annual temperature structure and its implications for heatwaves and monsoon variability.\\
In conclusion, this study demonstrates the utility of modern Bayesian approaches, specifically the coupled CTRW and the BSAR with isotonic GP models for understanding and forecasting the complex dynamics of temperature extremes in India. Their contrasting strengths offer a multifaceted view: one emphasizing irregular, memory-driven shifts and the other capturing smooth, monotonic changes. Together, these models enhance our understanding of climate change patterns and offer valuable tools for researchers, policymakers, and planners grappling with the increasing impacts of global warming in India.


\end{document}